\documentclass[12pt]{article}
\usepackage{graphics}

\newcommand{\lab}[1]{\label{#1}}

\newcommand{\ben}{\begin{enumerate}}
\newcommand{\een}{\end{enumerate}}
\newcommand{\be}{\begin{equation}}
\newcommand{\ee}{\end{equation}}
\newcommand{\bdm}{\begin{displaymath}}
\newcommand{\edm}{\end{displaymath}}
\newcommand{\bea}{\begin{eqnarray}}
\newcommand{\eea}{\end{eqnarray}}

\newcommand{\non}{\nonumber}
\newcommand{\real}{\mbox{Re\,}}

\newcommand{\pdone}[2]{\frac{\partial {#1}}{\partial {#2}}}
\newcommand{\sdone}[2]{\frac{
d {#1}}{d {#2}}}

\newcommand{\reseteqnos}[1]{\renewcommand{\theequation}
{#1.\arabic{equation}}\setcounter{equation}{0}}

\def\gsim{ \mbox{ \raisebox{-.9ex}{$\stackrel{\textstyle >}{\sim}$} } }
\def\lsim{ \mbox{ \raisebox{-.9ex}{$\stackrel{\textstyle <}{\sim}$} } }

\def\dsp{\displaystyle}

\newcommand{\up}[1]{$^{{#1}}$}

\title{Oscillations in a maturation model of blood cell production}
\author{Ivana Drobnjak and A.\,C.\ Fowler\\ Mathematical Institute,
Oxford University\\
24-29 St Giles', Oxford OX1 3LB, England\\ Michael C. Mackey\\
Departments of Physiology, Physics and Mathematics, and\\ Centre for
Nonlinear Dynamics,\\ McGill University, Montreal, Quebec\\ Canada}
\date{\today}

\begin{document}

\maketitle

\section{Introduction}
\reseteqnos{1}

A number of hematological diseases are characterised by
oscillations in the circulating density of various types of blood
cells. These include chronic myelogenous leukaemia (CML), cyclical
neutropenia (CN), polycythemia vera (PV) and aplastic anaemia
(AA). Examples of blood cell counts for CML and CN are shown in
figures \ref{figcml} and \ref{figcn}.

A review of the clinical data, and a discussion of possible
mechanisms for the oscillations, is given by Haurie {\it et al.}\
(1998). These mechanisms focus on the r\^ole of negative feedback
control on proliferation and differentiation of blood cells within
the bone marrow, together with time delays due to cell cycling and
maturation. There are consequently a number of different ways in
which oscillations can occur, and one object of mathematical
modelling of blood cell development is to understand which of
these effects may be responsible for the oscillations which are
seen.

Blood cells are produced  through a process of differentiation
from primitive stem cells in the bone marrow. These pluripotential
stem cells begin to develop along one of several different cell
lineages, forming blast cells which eventually develop through a
number of different stages to form the various different kinds of
blood cells. The most numerous are the red blood cells, or
erythrocytes, whose normal density in the blood is about $5\times
10^6$ cells $\mu$l\up{-1}. Their primary function is in
transporting oxygen to the tissues. Platelets are formed by the
fragmentation  of megakaryocytes, which develop in the bone
marrow. Platelets are present at levels of $5\times 10^5$ cells
$\mu$l\up{-1}, and their function is in blood clotting. Finally,
there are a number of different white blood cells, the most common
of which are neutrophils (5000 cells $\mu$l\up{-1}) and
lymphocytes (2000 cells $\mu$l\up{-1}), which form constituent
parts of the immune system. The normal levels of these cells are
controlled by a number of mechanisms, and excess or deficit of the
various cell types defines certain kinds of disease; for example,
anaemia refers to a low red blood cell count, below $4\times 10^6$
cells $\mu$l\up{-1}.

\begin{figure}
\begin{center}
\resizebox{280pt}{240pt}{\includegraphics{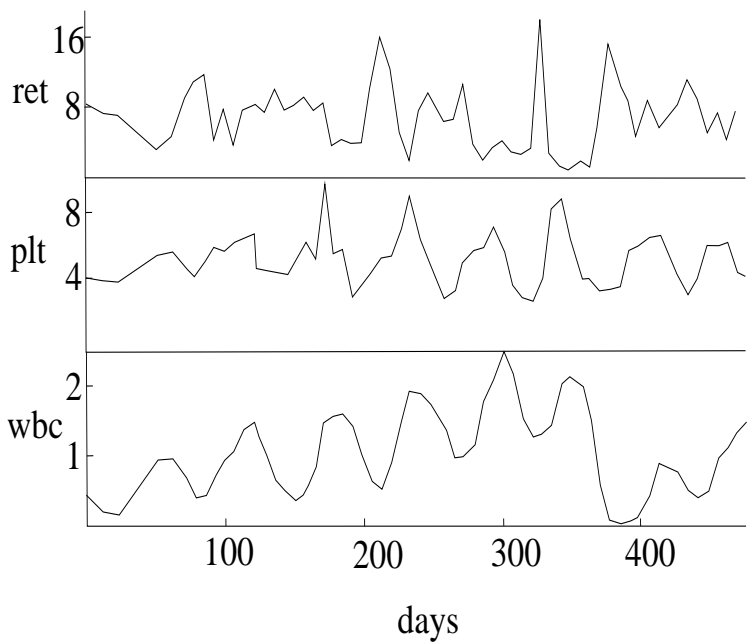}} \caption{\lab{figcml}Oscillations in white blood cell,
platelet and reticulocyte numbers in a patient with chronic myelogenous leukaemia. The units are white blood
cells, 10\up{5} cells $\mu$l\up{-1}; platelets, 10\up{5} cells $\mu$l\up{-1}; reticulocytes, 10\up{4} cells
$\mu$l\up{-1}. Redrawn from Chikkappa {\it et al.} (1976).}
\end{center}
\end{figure}

\begin{figure}
\begin{center}
\resizebox{220pt}{240pt}{\includegraphics{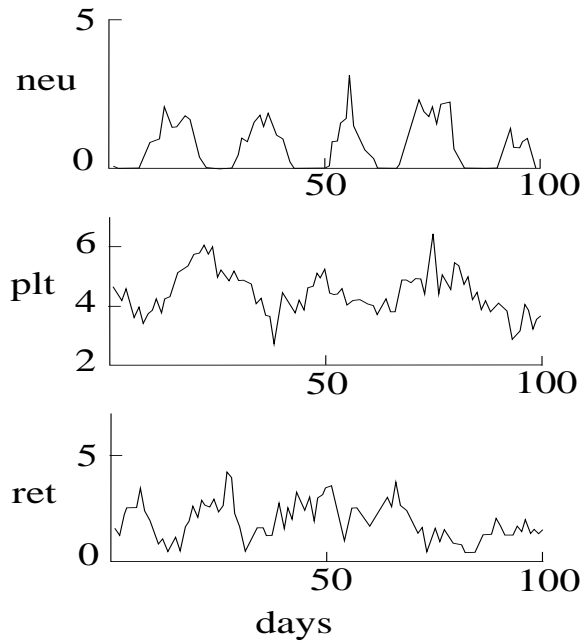}} \caption{\lab{figcn}Oscillations in neutrophils, platelet
and reticulocyte numbers in a patient with cyclical neutropenia. The units are neutrophils, 10\up{3} cells
$\mu$l\up{-1}; platelets, 10\up{5} cells $\mu$l\up{-1}; reticulocytes, 10\up{4} cells $\mu$l\up{-1}. Redrawn
from Haurie {\it et al.}\ (1998).}
\end{center}
\end{figure}

There are a number of features in figures \ref{figcml} and \ref{figcn}
which are of note. In CML, there are regular oscillations in white
blood cell counts with a long period ranging from 40 to 80 days
(Fortin and Mackey (1999)).
%There is
%also an even longer accumulation and subsequent collapse whic is
%evident in this figure.
The other cell lines (platelets and reticulocytes, i.\,e.\
erythrocyte precursors) also oscillate in a similar fashion (figure
\ref{figcml} does not show this well: see Fortin and Mackey (1999) for
other examples).

A similar observation is true of cyclical neutropenia. Oscillation
periods are of order 20 days, during which there is a marked
collapse of the neutrophil count to vanishingly low levels (Dale and
Hammond 1988, Guerry {\it et al}.\ 1973).
Other  cell types oscillate, but only the neutrophils appear to
oscillate fairly regularly: oscillations in other cell types (e.\,g.,
red blood cells, platelets, reticulocytes and lymphocytes) are
marked by irregularity and high frequency `noise' (Guerry {\it et
  al}.\ 1973). This latter feature is well illustrated in figure \ref{figcn}.

The purpose of the present paper is throw some light on these
observations by the study of a model of blood cell proliferation and
differentiation. This model is similar to those of previous authors,
particularly that of Mackey and Rudnicki (1994),
and describes the stem cell and developing (blast) cell
populations as functions of time, age (time through the proliferative
cell cycle), and maturation (stage in the differentiation
process). Fokas {\it et al.}\ (1991) describe a model with discrete
generations in the development of blast cells, while Mackey and
Rudnicki (1994) develop a corresponding continuous model (i.\,e.,
developmental stage is a continuous variable).

In this paper we use a continuous model to describe the
development of a single cell lineage following the committal of
stem cells. Three separate controls are implemented in the model,
namely the proliferative control of stem cells, the proliferative
control of developing blast cells, and the peripheral control of
stem cell committal by circulating blood cell density. We show
that variation of parameters in all three control systems can
cause oscillations, and that the characters of these oscillations
are very different. This allows us some potential insight into the
mechanisms that  may be operative in some of these dynamic blood
diseases.

\section{A model of maturation of blood cell production}
\reseteqnos{2}

The basic model is similar to that described by Mackey and
Rudnicki (1994).  It has been analysed in various versions by Rey
and Mackey (1993), Crabb et al.\ (1996) and Mackey and Rudnicki
(1999).  A particular feature of these models was the assumption
of zero maturation rate at maturation state zero.  %This leads to
%some odd behaviour, which we believe to arise because the model
%does not properly identify the r\^{o}le played by primitive stem
%cells.
In our formulation of the model, we do not make this assumption.

We consider all cell lineages to consist of populations of two
types, proliferative and resting phase.  These are denoted $p$ and
$n$, respectively, and are functions of age $a$ (time since the
inception of the proliferative cell cycle) and maturation $m$
(degree of maturation, measured in maturation units (mat), which
could be, for example, cell generation number). Also, $p$ and $n$
are functions of time $t$. Thus, we have $p=p(t,m,a)$ and
$n=n(t,m,a)$.  The dimensions of $p$ and $n$ are cells age$^{-1}$
mat$^{-1}$.

In the cell population, there will be a finite number which are
primitive and have not begun differentiation.  These can {\it not}
be characterised in terms of $p$ and $n$ at $m=0$, since the
latter
    are cell densities with respect to $a$ {\it and} $m$.  The
    primitive stem cells are  characterised by cell densities
    $p_0(t,a)$ and $n_0(t,a)$, such that $p_0\,da$ and $n_0\,da$
    are the numbers
    of primitive stem cells (with $m=0$) of age in $(a,a+da)$.

\begin{figure}[h]
\begin{center}
\resizebox{12cm}{12cm}{\includegraphics{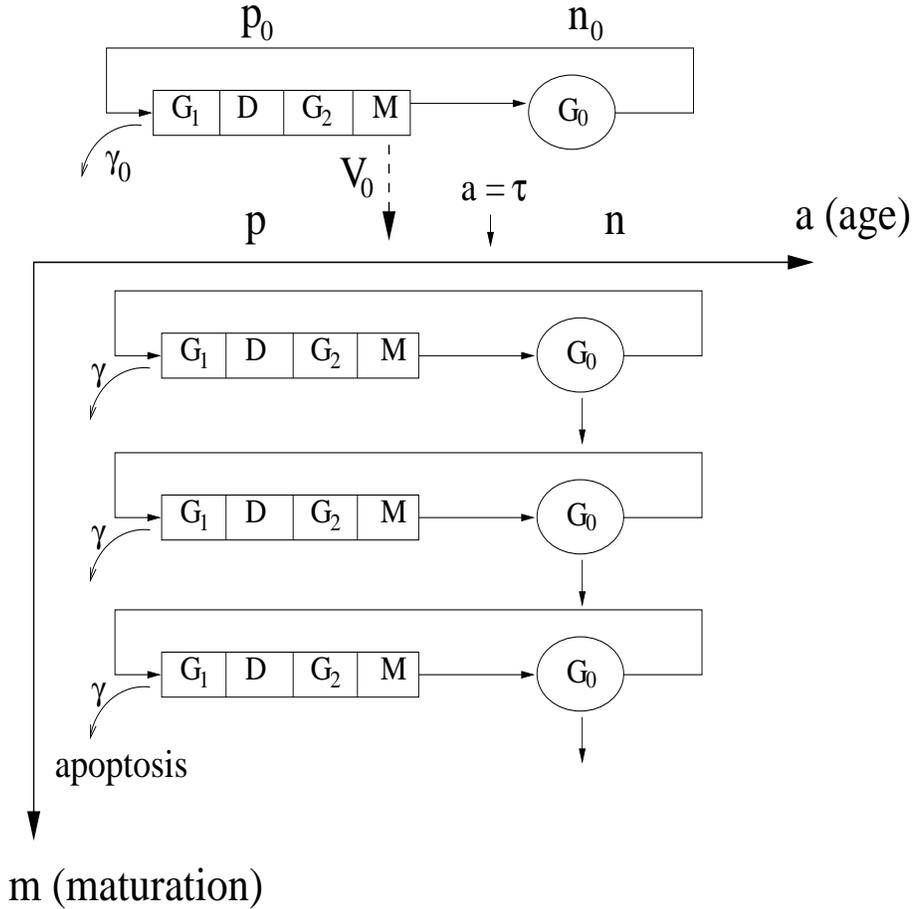}} \caption{Schematic evolution of cells. Each cell {\it ages}
as it goes through its cell cycle, before dividing and entering a resting ($G_0$) phase; at the same time, the
cells mature. The time-like variables $a$ (age) and $m$ (maturation) are independent.} \lab{fig1}
\end{center}
\end{figure}

The evolution of the system is illustrated schematically in figure
    \ref{fig1}.  We suppose that cell mortality occurs at a rate
    $\gamma$ (for proliferating cells only), and that cell
    maturation occurs continuously at a rate $V$ (for  both
    proliferative and resting phases).  We suppose that both
    $\gamma$ and $V$ may depend on maturation stage $m$, but not
    on $t$. Conservation of
    proliferative cells then implies
\be \frac{\partial
    p}{\partial t}+\frac{\partial p}{\partial a}+\frac{\partial
    (Vp)}{\partial m}=-\gamma p,
\lab{2.1}
\ee
where the units of
    $V$ are mat d$^{-1}$ (maturation units per day).  We suppose
    (\ref{2.1}) applies during
    a cycle of length $\tau$ (which might depend on $m$), thus for
    $0<a<\tau$; then for $a>\tau$, the cells in the resting phase
    satisfy the equation
\be \frac{\partial n}{\partial
    t}+\frac{\partial n}{\partial a}+\frac{\partial (Vn)}{\partial
    m}=-Rn,\lab{2.2}
\ee
which differs from (\ref{2.1}) by the
    rate of recruitment $R$ back to the proliferative phase;
    resting cell
    mortality is taken to be zero.  Equation (\ref{2.2}) applies
    for $a>\tau$.

At the end of the cell cycle, $a=\tau$, we apply a boundary condition
    describing the conversion of $p$ to $n$.  Mackey and Rudnicki
    (1994) allow a very general condition, on the basis that
    cells at maturation $M$ can divide to form cells at maturation
    $g(M)\leq M$.  Specifically, this implies $2p[t,M,\tau
    (M)]\,dM=n[t,g(M),\tau\{g(M)\}]\,dg$.  If we write $m=g(M)$,
    $M=h(m)$ (so $h=g^{-1}$), then this becomes \be n[t,m,\tau
    (m)]=2p[t,h(m),\tau\{h(m)\}]h'(m),\lab{2.3} \ee where
    $g(m)\leq m$ implies $h(m)\geq m$.

A boundary condition for $p$ at $a=0$ follows from the recruitment
    condition (the renewal equation)
\be
    p(t,m,0)=RN(t,m),\lab{2.4}
\ee
where we introduce the total
    resting cell density \be N=\int^\infty_\tau n\,da.\lab{2.5} \ee

Now we integrate (\ref{2.2}) from $a=\tau$ to $a=\infty$, taking
    $n\rightarrow 0$ as $a\rightarrow\infty$ (which is necessary
    if there are a finite number of cells).  We suppose that
    $V=V(m)$ is independent of $a$ and $t$, and
    and $R=R(t,m)$ is independent of $a$.  Then
\be \frac{\partial N}{\partial t}+\frac{\partial
    (VN)}{\partial m}=-RN+2p[t,h(m),\tau\{h(m)\}]h'(m),\lab{2.6}
    \ee
adopting (\ref{2.3}).

We need to solve (\ref{2.1}) for $p$.  We use the method of
    characteristics, and begin by applying the recruitment condition
    (\ref{2.4}). Specifically, we apply the parametric conditions
\be\lab{2.6.1}t=s,\quad m=\mu,\quad a=0,\quad p=R(s,\mu)N(s,\mu),\ee
valid for $s,\mu >0$;
    then the characteristic solution is
\begin{eqnarray} a&=&t-s,\
    \ \int^m_\mu\frac{d\rho}{V(\rho)}=t-s,\nonumber\\
    p&=&R(s,\mu)N(s,\mu)\exp\left[-\int^t_s[\gamma
    +V'(m)]\,dt\right].\lab{2.7} \end{eqnarray}

Define a function $\nu (m,a)$ by
\be
    \int^m_\nu\frac{d\rho}{V(\rho)}=a.\lab{2.8}
\ee
Then
    $a=t-s$, $\mu=\nu (m,a)$.  Also $dt=dm/V(m)$ on a
    characteristic, thus for $t>a$ (and also $\nu >0$),
\be
    p(t,m,a)=R[t-a,\nu (m,a)]N[t-a,\nu
    (m,a)]\exp\left[-\int^m_{\nu (m,a)}\{\gamma
    +V'(\rho)\}\frac{d\rho}{V(\rho)}\right];\lab{2.9}
\ee
    simplifying and putting $a=\tau$, we have
\be
    p(t,m,\tau)=R[t-\tau,\nu (m,\tau)]N[t-\tau,\nu (m,\tau)]
    \exp\left[-\int^m_{\nu (m,\tau)}\frac{\gamma
    \,d\rho}{V(\rho)}\right]\frac{V[\nu
    (m,\tau)]}{V(m)},\lab{2.10} \ee
for $t>\tau$ and $\nu >0$.
Finally, (\ref{2.6}) becomes
\[ \hspace*{-2.5in}
\frac{\partial N}{\partial
    t}+\frac{\partial}{\partial m}(VN) =-RN\]
\[\hspace*{0.5in}
+2h'(m)R[t-\tau,\nu\{h(m),\tau\}]N[t-\tau,\nu\{h(m),\tau\}]\times\]
\be\hspace*{2in}
\exp\left[-\int^{h(m)}_{\nu\{h(m),\tau\}}\frac{\gamma\,
    d\rho}{V(\rho)}\right]\frac{V[\nu\{h(m),\tau\}]}{V[h(m)]}.\lab{2.11}
    \ee
Note that
\be \int^m_{\nu
    (m,\tau)}\frac{d\rho}{V(\rho)}\equiv\tau.\lab{2.12} \ee

It is convenient to define a modified maturation variable $\xi$  by
    \be \xi=\int^m_{0}\frac{d\rho}{V(\rho)};\lab{2.13} \ee
    $\xi$ has units of time, and indeed it is equal to the elapsed
    time during maturation. Note that $\nu >0$ if $\xi >\tau$.
The lower limit can be chosen for convenience, and allows us to fix $\xi$ at
    some reference point; here we choose this to be the initial
    maturation stage (note that this cannot be done if $V(0)=0$).  Define
    also
\be \eta (\xi)=\int^{h(m)}_{0}\frac{d\rho}{V(\rho)}
    \lab{2.14}\ee (note $\eta\geq\xi$ since $h\geq m$).  Now if
\be
    F(m)\equiv f(\xi),\lab{2.15} \ee then we find
\bea
    F[h(m)]&=&f(\eta),\non\\
    F[\nu\{h(m),\tau\}]&=&f(\eta-\tau).\lab{2.16} \eea
    We change variable from $m$ to $\xi$, and define
    \bea v(\xi)&\equiv &V(m),\non\\ M&\equiv &
    NV\lab{2.17} \eea
(note that $Md\xi=Ndm$, so that
    $M$ is cell density in terms of the variable $\xi$; the units
    of $M$ are cells d$^{-1}$).  After a
    little manipulation, we find
\be\lab{2.18.2} \frac{\partial
    M}{\partial t}+\frac{\partial M}{\partial \xi}=-RM+Q,\ee
where
\be
    Q=2\eta'(\xi)R[t-\tau,\eta-\tau]
    M[t-\tau,\eta-\tau]\exp\left[-\int^\eta_{\eta-\tau}\gamma\,
    d\xi\right],\lab{2.18} \ee
where we write
    $\gamma$, $R$ and $M$ as dependent on $\xi$ rather than
    $m$. This equation applies if $t>\tau$ and $\eta >\tau$.

In order to find the form of the source term for $t<\tau$ or $\eta
<\tau$, we must solve the equation (\ref{2.1}) for $p$ using the
initial data from $m=0$ and $t=0$. If, specifically, we have an
initial condition
\be\lab{2.18.1}p=p_I(m,a)\quad {\rm at}\quad t =0,\ee
then
after some algebra
we find that
\be\lab{2.18.3}Q=
2\eta'(\xi)p_I[\eta-t,\tau-t]v(\eta-t)
\exp\left[-\displaystyle\int^{\eta}_{\eta -t}\gamma\, d\xi\right],
\quad t<\tau,\quad \eta >t.
\ee

The definition of $Q$ in $t>\eta$ and $\eta<\tau$ requires
consideration of the stem cell evolution, and we now turn to this.
    Conservation laws for the stem cell  densities $p_0$ and $n_0$
are
\begin{eqnarray}
    \frac{\partial p_0}{\partial t}+\frac{\partial p_0}{\partial
    a}&=&-(\gamma_0+V_0)p_0, \nonumber\\ \frac{\partial
    n_0}{\partial t}+\frac{\partial n_0}{\partial
    a}&=&-(V_0+R_0)n_0,\lab{2.20} \end{eqnarray}
where $V_0$ is
    the  rate of loss of stem cells to maturation, $R_0$ is the
    stem cell recruitment rate from the resting phase, and $\gamma_0$
    is the mortality rate of stem cells in the proliferative
    phase.  We allow $R_0$, $V_0$ and $\gamma_0$ to depend on $t$,
    but we assume they are independent of $a$. Note that $V_0\neq
    0$, indeed $V_0\neq V(0)$, as the units of $V_0$ and $V$ are
    not even the same: $V$ has units of mat d$^{-1}$, while $V_0$
    has units of d$^{-1}$.  Note also that $p_0$ and $n_0$ have
    units of cells age$^{-1}$ (unlike $p$ and $n$).

The primitive loss to maturation must balance the source for $p$ and
    $n$ at $m=0$, thus
\be V_0p_0=(Vp)|^{}_{m=0},\quad
    V_0n_0=(Vn)|^{}_{m=0},\lab{2.21}\ee
and the units are
    consistent.

Analogously to (\ref{2.4}) and (\ref{2.3}), we have
    \bea p_0(t,0)&=&R_0(t)N_0(t),\non\\
    n_0(t,\tau)&=&2p_0(t,\tau), \lab{2.22} \eea
where
    \be N_0=\int^\infty_\tau n_0\,da.\lab{2.23} \ee
Integration
    over $a$ now yields
\be
    \frac{dN_0}{dt}=-V_0N_0-R_0N_0+2p_0|^{}_{a=\tau},\lab{2.24}
    \ee
and
\be (NV)|^{}_{m=0}=N_0V_0.\lab{2.25}
\ee

In order to find $p_0$ we must solve
\be \frac{\partial p_0}{\partial
    t}+\frac{\partial p_0}{\partial
    a}=-(\gamma_0+V_0)p_0,\lab{2.26} \ee
with parametric initial
    conditions:
\[p_0=p_{00}(\alpha),\quad a=\alpha>0,\quad t=0,\]
\be p_0=R_0(s)N_0(s), \quad a=0, \quad
    t=s>0.\lab{2.27} \ee
For $t> a$, the solution is
\be
    p_0=R_0(t-a)N_0(t-a)\exp\left[-\int^t_{t-a}
    [\gamma_0(t')+V_0(t')]\,dt'\right],\lab{2.28}
    \ee
whereas for $t<a$,
\be\lab{2.28.1}p_0=p_{00}(a-t)
    \exp\left[-\int^t_{0}[\gamma_0(t')+V_0(t')]\,dt'\right].\ee
Putting $a=\tau$, we find
\be
    \frac{dN_0}{dt}=-(R_0+V_0)N_0+2R_0(t-\tau)N_0(t-\tau)
\exp\left[-\int^t_{t-\tau}
    [\gamma_0(t')+V_0(t')]\,dt'\right]
,\quad t>\tau,
    \lab{2.29} \ee
which prescribes the control system for
    $N_0$, analogously to that of Mackey (1978).  For
$t<\tau$, the equation for $N_0$ involves the initial condition for
$p_0$, and we can equivalently simply prescribe initial data for $N_0$
there.

Finally, we complete the definition of $Q$ in (\ref{2.18.2}) by
solving (\ref{2.1}) using the initial data on $m=0$:
\be\lab{2.29.1}m=0,\quad a=\alpha>0,\quad t=s>0,\quad
V(0)p=V_0(s)p_0(s,\alpha).\ee
We find
\be\lab{2.29.2}p(t,\xi,a)=\frac{V_0(t-\xi)p_0(t-\xi,a-\xi)}{v(\xi)}\exp\left[-\int_0^\xi\gamma\,d\xi\right],\quad
\xi<a,\quad \xi<t,\ee
from which it follows that
\be\lab{2.29.3}Q=2\eta'(\xi)V_0(t-\eta)p_0[t-\eta,\tau-\eta]
\exp\left[-\displaystyle\int^{\eta}_{0}\gamma\, d\xi\right],
\quad t>\eta,\quad \eta <\tau.\ee
Along with (\ref{2.18}) and (\ref{2.18.3}), this completes the
definition of $Q$ for $t>0$, $\eta>0$ (and thus $\xi>0$). In summary,
\be\lab{2.29.4}Q=\left\{\begin{array}{ll}
2\eta'(\xi)R[t-\tau,\eta-\tau]M[t-\tau,\eta-\tau]
\exp\left[-\displaystyle\int^\eta_{\eta-\tau}\gamma\, d\xi\right],&
\quad t>\tau,\quad \eta >\tau,\\
&\\
2\eta'(\xi)p_I[\eta-t,\tau-t]v(\eta-t)
\exp\left[-\displaystyle\int^{\eta}_{\eta -t}\gamma\, d\xi\right],&
\quad t<\tau,\quad \eta >t,\\
&\\
2\eta'(\xi)V_0(t-\eta)p_0[t-\eta,\tau-\eta]
\exp\left[-\displaystyle\int^{\eta}_{0}\gamma\, d\xi\right],&
\quad t>\eta,\quad \eta <\tau.\\
\end{array}\right.
\ee

The two
    equations (\ref{2.29}) and (\ref{2.18.2}) are coupled through
    (\ref{2.25}), which provides the requisite boundary condition
    for $M$ at $\xi=0$:
\be
    M=V_0N_0 \ \ \mbox{at} \ \ \xi =0.\lab{2.30} \ee
We see that
    by an appropriate consideration of the primitive stem cells, we derive
    a coherent model which does not require $V(0)=0$.

Many authors (for example, see Rey and Mackey (1993) and Dyson et
    al.\ (1996)) study the differential equation (\ref{2.11}) for
    $N$, under the assumption that $V$ does tend to zero as
    $m\rightarrow 0$, for example,
\be V=rm.\lab{2.19} \ee
The
    reasoning behind this is that, if primitive stem cells mature
    at a finite rate, then such cells will be immediately lost to
    $m>0$, which makes no physiological sense, because the cell
    population then inexorably disappears.  Only by choosing
    $V(0)=0$ can we allow primitive stem cells to endure.
In the present version of the model, it
    is also possible to allow $V(0)=0$, for example, (\ref{2.19})
    would then imply
\be M\rightarrow 0 \ \ \mbox{as} \ \
    \xi\rightarrow-\infty.\lab{2.31} \ee
The sensitivity of the solution to this condition has led to the
idea that the system may have unstable and even chaotic solutions
(e.g., Crabb et al.\ (1996)), because of the degeneracy of the
equation at $m=0$. Our considerations here suggest that the
requirement that $V(0)=0$ is  inaccurate, because it does
not properly address the biological question of how the primitive
stem cells should be described.

\section{Dimensionless model}
\reseteqnos{3}

How we solve the model depends on the complexity of our assumptions
    about $\gamma$, $R$, $\eta$ and $R_0$.  In the remainder of this
    paper we will assume $g(m)=m$, thus $\eta =\xi$, and that
    $\gamma$ and $\gamma_0$ are constant.  The equation for the
    maturing cells, (\ref{2.18.2}), is
\be\lab{3.01}\frac{\partial
M}{\partial t}+\frac{\partial M}{\partial \xi}=-RM+Q,\ee
and is a hyperbolic delay-partial differential equation. Figure
\ref{fig1.1} shows the regions where the different definitions
of $Q$ apply. In regions II and segment ({\it a}) of region III, that
is for $t<\tau$ and all $\eta =\xi>0$, $Q$ depends on the initial
    data, either $p_I$ (in II) or $p_{00}$ (in III({\it a})). Thus
    we may equivalently simply choose instead to prescribe $M$ in
    $0<t<\tau$, and this we do. In fact, since $\xi$ is finite,
    the part of the solution which depends on this initial data
    will wash out of the system in a finite time. It is therefore
apparently of little concern.

\begin{figure}[t]
\begin{center}
\resizebox{240pt}{240pt}{\includegraphics{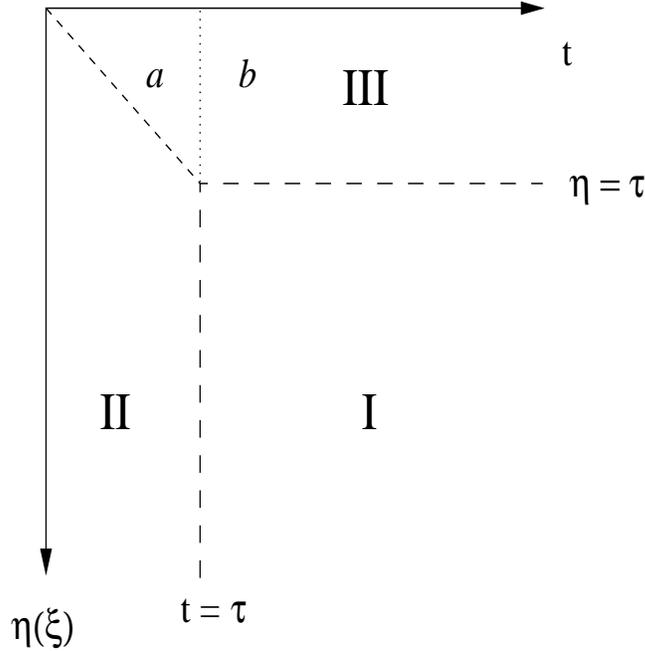}} \caption{\lab{fig1.1}Regions of different definitions of
$Q$ in (\ref{2.29.4}). Regions I, II and III correspond to the first, second and third definitions of $Q$ and
their locations of validity in the $(t,\eta)$ plane. The vertical dotted line in region III (where $Q$ is
defined in terms of $p_0$) separates the region ({\it a}) where (\ref{2.28.1}) applies (to the left) from that
({\it b}) where (\ref{2.28}) applies (to the right).}
\end{center}
\end{figure}

We therefore confine ourselves to consideration of the definition of
$Q$ in regions I and III({\it b}); with the assumptions we have made,
we find that for $t>\tau$,
\be\lab{3.03}Q=\left\{\begin{array}{ll}
2e^{-\gamma\tau}R[t-\tau,\xi-\tau]M[t-\tau,\xi-\tau],&
\quad \xi >\tau,\\
&\\
2e^{-\gamma_0(\tau-\xi)}e^{-\gamma\xi}
\exp\left[\dsp{-\int_{t-\tau}^{t-\xi}V_0(t')\,dt'}\right]
V_0(t-\xi)R_0(t-\tau)N_0(t-\tau),& \quad \xi <\tau.\\
\end{array}\right.
\ee
We thus have to solve (\ref{3.01}) with (\ref{3.03}) in $t>\tau$, with the
boundary condition (\ref{2.30}) on $\xi = 0$, and prescription of an
initial function for $M$ in $0<t<\tau$.

A principal
issue of focus is  how the recruitment rates $R$ and
    $R_0$ and the committal rate $V_0$ depend on $M$, $N_0$ and
$\xi$.  There is very little to
     constrain our choice.
In what follows, we assume $R_0=R_0(N_0)$ (stem cell proliferation is
controlled by stem cell density).
We follow   Mackey and Rudnicki (1994) in supposing that $R$ depends on the
    total differentiating cell population $\bar{M}$, where
\be \bar{M}(t)=\int_0^{\xi_F}M\,d\xi,\lab{3.2} \ee
$\xi_F$ being the time of final maturation,
\be\lab{3.2.1}\xi_F=\int_0^{m_F}\frac{d\rho}{V(\rho)},\ee
and $m=m_F$ at full maturity.
We suppose  that the rate of committal $V_0$ should
    depend on the peripheral blood cell count, $B$, thus
$V_0=V_0(B)$. A simple model
    for $B$ is
\be\lab{3.2.2}\sdone{B}{t}=M|_{\xi_F}-\gamma_BB,\ee
where $\gamma_B$ is the specific decay rate of the peripheral blood
    cells, and the source term $M|_{\xi_F}$ is the delivery rate
    to the blood from the maturation phase cells.
Peripheral control models of similar
    type have been studied by Bernard {\it et al.}\ (2003).
Assumptions of this type  are liable to be important in the
    evolution of diseases such as cyclical neutropenia, which is
thought to be due to an instability in the peripheral control of stem
cell committal.  In addition, it is likely
    that other controls affect rate of apoptosis, maturation rate, cell
    cycle time, and so on.

The equation for $N_0$ (\ref{2.29}) now takes the form
\be
\dot{N}=-[V_0+R_0(N_0)]N_0+2e^{-\gamma_0\tau}R_0(N_{0\tau})N_{0\tau}
\exp\left[-\int^t_{t-\tau}V_0[B(t')]\,dt'\right],\lab{3.3}\ee
where $N_{0\tau}=N_0(t-\tau)$.
    This is precisely the model of Mackey (1978) if $V_0$ is constant, and
    has been studied by Fowler and Mackey (2002) in the limit
\be V_0\tau\ll 1, \lab{3.5} \ee
when it is shown
    that relaxation oscillations will occur for a further
    parameter $\mu_0=[2e^{-(\gamma_0+V_0)\tau}-1]/V_0\tau$ within
    a certain $O(1)$ range.
    (Note that in the notation of Fowler and Mackey's model,
    $\gamma =\gamma_0+V_0$, $\delta =V_0$.)  When such
    oscillations occur, they will propagate through the maturing
    cells; however we show in this paper that the resultant
amplitude of oscillations of mature blood cells is small unless
amplification also occurs during maturation.

We can write (\ref{3.03}) in abbreviated form as
\be\lab{3.6}Q=\left\{\begin{array}{ll}
2e^{-\gamma\tau}
R(\bar{M}_\tau)M_{\tau,\tau},&
\quad \xi >\tau,\\
&\\
2e^{-\gamma_0(\tau-\xi)}e^{-\gamma\xi}
\exp\left[-\dsp{\int^{t-\xi}_{t-\tau}V_0[B(t')]\,dt'}\right]
V_0[B(t-\xi)]R_0(N_{0\tau})N_{0\tau},& \quad \xi <\tau,\\
\end{array}\right.
\ee
where $\bar{M}_\tau=\bar{M}(t-\tau)$, and
    $M_{\tau,\tau}=M(t-\tau,\xi-\tau)$.

We non-dimensionalise the model  by following the analysis of
    Fowler and Mackey (2002), which motivates a choice of  scales
    for  the variables as follows:
\[t, \xi\sim \tau,\quad M\sim M^*,\quad
    N_0=N_0^*S,\quad Q\sim\frac{M^*}{\tau},\quad \]
    \be\lab{3.7}
R_0=R_0^*h_0,\quad
    R= R^*h,\quad V_0=V_0^*v_0,\quad B\sim\frac{M^*}{\gamma_B}, \ee
where $R_0^*,\ R^*,\ V_0^*,\ N_0^*,\ M^*$ are
    determined by the  control functions (so that they are
    $O(1)$ functions of $O(1)$ variables).  For example,
    Mackey (1978) chooses for $R_0$ the Hill function
\be
    R_0=\frac{R_0^*}{1+(N_0/\theta)^{^{\scriptstyle n}}}.\lab{3.8} \ee
In this case, we would choose $N_0^*=\theta$ and $h_0$ is the Hill function
\be h_0(S)=\frac{1}{1+S^{^{\scriptstyle n}}}. \lab{3.9} \ee

The dimensionless stem cell equation is
\be
    \dot{S}=b_0\left[(1+\lambda_0)h_0(S_1)S_1
\exp\left(\varepsilon_0\left\{1-\int^t_{t-1}v_0[B(t')]\,dt'\right\}\right)
-h_0S\right]-\varepsilon_0v_0S,\lab{3.13} \ee
where
\be
    \varepsilon_0=
V_0^*\tau,\quad\lambda_0=2e^{-(\gamma_0+V_0^*)\tau}-1,\quad
    b_0=R_0^*\tau. \lab{3.14}
    \ee

The
    dimensionless form of (\ref{3.6}) is
\be\lab{3.10}
    \frac{\partial M}{\partial t}+\frac{\partial
    M}{\partial\xi}=-bh(\bar{M})M+Q,\ee
where
\be\lab{3.10.1}
Q=
\left\{\begin{array}{ll}
&b(1+\lambda)h(\bar{M}_1)M_{1,1}\,,
\quad \xi >1,\\
&\\
&\nu b_0(1+\lambda_0)e^{-\alpha\xi}
\exp\left[\varepsilon_0\left\{1-\dsp{\int^{t-\xi}_{t-1}v_0[B(t')]\,dt'}
\right\}\right]
v_0[B(t-\xi)]h_0(S_1)S_1, \quad \xi <1,\\
\end{array}\right.\ee
in which
\be \nu =\frac{N_0^*V_0^*}{M^*},\quad
b=R^*\tau, \quad \lambda
    =2e^{-\gamma\tau}-1,\quad\alpha=(\gamma-\gamma_0)\tau.\lab{3.11} \ee
This is
    analogous to the scaling used by Fowler and Mackey (2002).
The boundary condition for $M$ is
\be
    M=\nu v_0S \quad \mbox{at} \quad \xi=0,\lab{3.15} \ee
and if
\be\lab{3.15.1}M=M_f\quad \mbox{at} \quad \xi=\xi_f,\ee
then
\be\lab{3.15.2}\delta \dot{B}=M_f-B,\ee
where
\be\lab{3.15.3}\delta=\frac{1}{\gamma_B\tau},\quad
\xi_f=\frac{\xi_F}{\tau}.\ee
This completes the statement of the dimensionless form of the model.

\subsection*{Parameter values}

The equation (\ref{3.13}) is exactly that studied by Fowler and Mackey
(2002) (if $v_0\equiv 1$). However, their model can also   be
interpreted as a
`lumped', or compartmentalised, version of (\ref{3.10}) for the
maturing cells. One way of enabling  this is if we make the special
assumption that the maturation rate $V \to 0$ as both $m\to 0$ and
$m\to m_F$, as also assumed by Mackey and Rudnicki (1994). In this
case the range of $\xi$ is $(-\infty, \infty)$, and we have $M\to 0$
at both limits. Then integration of (\ref{3.10}) over $\xi$ again
leads to an equation of the form of (\ref{3.13}). In the present paper
we assume $V$ is finite at $m=m_F$, i.\,e., the mature blood cells are
delivered to the bloodstream at a finite rate, and this is then the
essential difference between the models with and without maturation.

In estimating the parameters, we follow Fowler and Mackey (2002)
in choosing $\tau\sim 2$ d, and we suppose that proliferative
control is effected at typical rates $R_{}^*\sim R_0^*\sim 2$
d\up{-1}. We suppose apoptosis rates are of order
$\gamma\sim\gamma_0\sim
 0.2$ d\up{-1}, and that committal rates are of order $V^{*}_0\sim 0.05$
d\up{-1}, and from these we find
\be\lab{3.15.4}
b\sim b_0\sim 4,\quad \lambda\sim \lambda_0\sim 0.3,\quad
\varepsilon_0\sim 0.1.\ee
The parameter $\alpha$ is not independent of the others, as
\be\lab{3.15.5}\alpha=\varepsilon_0+\ln\left(\frac{1+\lambda_0}{1+\lambda}
\right),\ee
and plausibly $\alpha\approx \varepsilon_0$.

The remaining parameters are $\delta$, $\nu$ and $\xi_f$. For $\delta$, we
assume a half life ($\gamma_B^{-1}$) of 7 hours, appropriate for
neutrophils (but, for example, certainly not for erythrocytes); then
\be\lab{3.15.6}\delta\sim 0.15.\ee

We can get some sense of the size of the remaining parameters
$\nu$ and $\xi_f$ by considering the nature of stem cells. These
are difficult to isolate; indeed it is not yet clear whether
genuine stem cells have ever really been isolated. The reason for
this is that there are few of them, and maturing cells will
typically undergo about (or at least) twenty divisions before
emerging as mature blood cells. A typical numerical estimate for
the total number of blast cells is 10$^{12}$ per kg body weight,
while for stem cells, a corresponding estimate is 10$^{6}$
(Bernard {\it et al.} (2003), Mackey (2001)). If this is the case,
then it successively implies that the parameter $\nu$ in
(\ref{3.11}) is very small ($\approx 10^{-6}$), and therefore also
that the maturation time is long. Typical estimates of
$\xi_F\approx 10$--$20$ days are consistent with values of
$\xi_f\approx 5$--$10$, and in fact the small parameter
$\xi_f^{-1}$ then plays the r\^ole corresponding to that of the
small parameter $\varepsilon$ in Fowler and Mackey's (2002)
analysis.

\subsection*{Steady state}

To elaborate this discussion, we now describe the steady state. For
simplicity, we ignore the distinct definition of $Q$ in $\xi<1$, and
extend the definition in $\xi>1$ back to $\xi=0$. The steady solution
of (\ref{3.10}) and (\ref{3.10.1}) is, with $v_0=S=1$,
\be\lab{3.16}M=\nu e^{s\xi},\ee
where $s$ is the unique positive solution of the pair
\[s=bh(\bar{M})\left[(1+\lambda)e^{-s}-1\right],\]
\be\lab{3.17}\bar{M}=\frac{\nu}{s}\left(e^{s\xi_f}-1\right).\ee
(Uniqueness follows from the fact that $\bar{M}$ is monotonically
increasing with $s$, hence $h(\bar{M})$ is monotonically decreasing with $s$,
hence $bh(\bar{M})\left[(1+\lambda)e^{-s}-1\right]$ is monotonically
decreasing with $s$, while evidently $s$ is increasing.) We can see that
$s<\ln(1+\lambda)$, and $s$ will be close to this value if $b$ is
large. Note also that by choosing $\lambda_0$ and $\varepsilon_0$ to
have certain specific values which depend on $\lambda$, $b$, $\nu$ and
$\xi_f$, this solution consistently extends back to $\xi =0$, even
allowing for the distinct definition of $Q$ in $\xi<1$.

Numerical solutions do confirm the exponential variation of $M$ with
$\xi$. In general, it is found that $M$ decreases for $0<\xi<1$,
before subsequently increasing.

\section{Periodic solutions}
\reseteqnos{4}

We are interested in finding whether periodic solutions can
occur. There are three different controllers in the model, and thus
three different ways in which oscillations can occur: these are
described below. We define a default reference set of parameters, and
these are given in table \ref{table1}. They are those suggested by
independent estimate, except that we take $\nu=10^{-2}$ rather than
10\up{-6}. This is partly for numerical expediency, as smaller values of
$\nu$ require larger $\xi_f$ and thus longer computation times, and
also because the value of $\nu$ is not well constrained.

\begin{table}[h]
\begin{center}
\begin{tabular}{|c|c|} \hline
Symbol          & Typical value \\ \hline
$n$         & 3 \\
$\varepsilon_0$     &0.1    \\
$\lambda_0$     &0.3    \\
$b_0$           &4  \\
$\nu$                   &10\up{-2}  \\
$b$         &4  \\
$\lambda$       &0.3    \\
$\xi_f$         &5  \\
$\delta$            &0.2    \\
$v^*$                   &1  \\
$v'$                    &1  \\
\hline

\end{tabular}
\caption{\lab{table1}Default parameter values.}
\end{center}
\end{table}

We choose the Hill function controller (\ref{3.9}) for both
$h$ and $h_0$, thus
\be\lab{4.01}h(\bar{M})=\frac{1}{1+\bar{M}_{}^{^{\scriptstyle n}}},\quad
h_0(S)=\frac{1}{1+S_{}^{^{\scriptstyle n}}},\ee
and we take the peripheral controller $v_0$ to have the form
\be\lab{4.4a}v_0=[v^*-v'B]_+,\ee
with default values of the amplitude and slope parameters
to be $v^*=v'=1$. The choice of a threshold in (\ref{4.4a}) is
motivated by the observation that neutrophil populations can dwindle
to zero in cyclical neutropenia, which would appear to require zero
production for sufficiently high blood cell counts.

With this choice of the controller functions and using the default
parameters, the steady state is stable. Instabilities arising from
parameter variations are described below.

\subsection*{Numerical method}

We have to solve the two ordinary differential equations (\ref{3.13})
and (\ref{3.15.2}), and the partial differential equation
(\ref{3.10}). The solution is complicated by the presence of the
integral in (\ref{3.13}). We define
\be\lab{4.1}U=S\exp\left[\varepsilon_0\int^t v_0[B(t')]\,dt'\right],\ee
and then $S$ and $U$ satisfy the pair of equations
\[\dot{S}=S\left[\frac{\dot{U}}{U}-\varepsilon_0v_0\right],\]
\be\lab{4.2}\dot{U}=
b_0\left[(1+\lambda)e^{\alpha}h_0(S_1)U_1-h_0(S)U\right].\ee
On the assumption that $S$ remains bounded, $U$ grows exponentially as
$U\sim\exp(\overline{v}_0t)$, where $\overline{v}_0$ is the mean of
$v_0$. This is likely to cause difficulty in numerical solutions, and
this can be reduced by using the algebraically growing function $W=\ln
U$, whence
\[\dot{S}=S\left[\dot{W}-\varepsilon_0v_0\right],\]
\be\lab{4.3}\dot{W}=
b_0\left[(1+\lambda)h_0(S_1)e^{\alpha+W_1-W}-h_0(S)\right].\ee
In our numerical solutions, we solve (\ref{4.3}) using the second
order accurate improved Euler method, and we similarly solve
(\ref{3.10}) along the
characteristics $\xi-t=\eta$, on which the function $Q$ takes the form
\be\lab{4.4}
Q=
\left\{\begin{array}{ll}
&b(1+\lambda)h(\bar{M}_1)M_{1}\,,
\quad \xi >1,\\
&\\
& b_0(1+\lambda)e^{\alpha(1-\xi)}
M_0(\eta)h_0(S_1)\exp\left[W_1-W(\eta)\right], \quad \xi <1,\\
\end{array}\right.\ee
where $M_0(\eta)=M(\eta,0)$ and $M_1=M_{1,1}$, i.\,e., $M$ delayed by
one along the characteristic.

Accurate solutions are obtained with a time step $\Delta t=\Delta \xi =
0.05$, and these were checked aginst values $\Delta t=\Delta \xi =
0.02$ (which are used to give the figures).

\subsection*{Stem cell oscillations}

Oscillations in the primitive stem cell population will occur for a
finite range of the parameter $\lambda_0/\varepsilon_0$, as described
by Fowler and Mackey (2002), when $v_0=1$. For the default values of
$b_0=4$, $n=3$, the approximate range of instability is
$0.5\varepsilon_0\lsim \lambda_0\lsim 1.5\varepsilon_0$, and this is
modified in an obvious way when the peripheral controller alters the
value of $v_0$. Figure \ref{fig3} shows the oscillations which occur
in the stem cell population when $\lambda_0$ is reduced to 0.05. It is
an interesting fact that these oscillations are hardly manifested in
the blood cell population. The apparent reason for this is that the
small value of $\nu$ means that oscillations in $M_0$ are small, and
therefore also in $M_f$, because small perturbations propagate stably
down the maturation axis. The blood cell population is therefore
stable, and $B\approx M_f$.

\begin{figure}
\begin{center}
\resizebox{240pt}{200pt}{\includegraphics{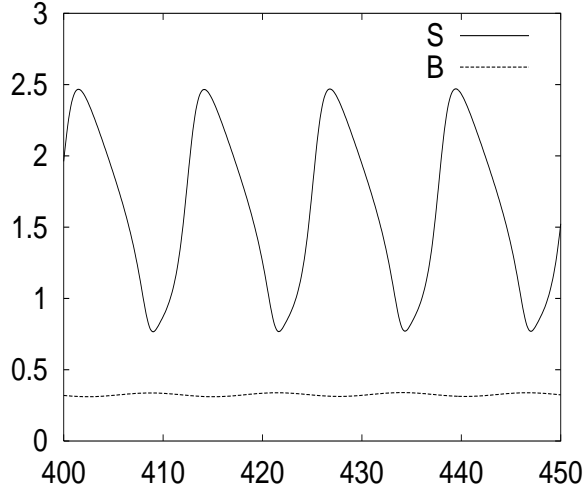}} \caption{\lab{fig3}Default parameter values, except that
$\lambda_0=0.05$. Stem cell oscillations are induced, without any significant effect on blood cells.}
\end{center}
\end{figure}

\subsection*{Proliferation-controlled oscillations}

We use the term proliferation-controlled oscillations to refer to
oscillations induced by destabilisation of the proliferative feedback
control function $h(\bar{M})$. If we compare the stem cell model
(\ref{3.13}) (with $v_0=1$)
\be\lab{4.5}
    \dot{S}=b_0\left[(1+\lambda_0)h_0(S_1)S_1
-h_0(S)S\right]-\varepsilon_0S \ee
with the blast cell model (along the characteristics)
\be\lab{4.6}\dot{M}=b\left[(1+\lambda)h(\bar{M}_1)M_{1,1}-
    h(\bar{M})M\right],\ee
it is not difficult to sense that modification of the parameters $b$
    or $\lambda$ may cause the blast maturation to proceed
    unstably.

This is what we find if $\lambda$ is increased to 0.6, and the
consequent oscillations are shown in figure \ref{fig4}. The steady
exponential proliferation of blast cells is unstable, and this causes
oscillations to occur in the maturation profile, and these
oscillations propagate along the characteristics, as shown in figure
\ref{fig5}.

\begin{figure}
\begin{center}
\resizebox{240pt}{200pt}{\includegraphics{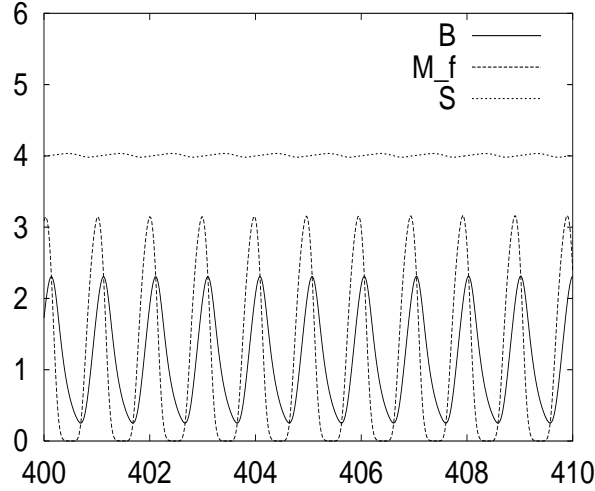}} \caption{\lab{fig4}Proliferatively controlled
oscillations due to increased proliferation. Default parameter values are used, except that $\lambda=0.6$.}
\end{center}
\end{figure}

\begin{figure}
\begin{center}
\resizebox{260pt}{200pt}{\includegraphics{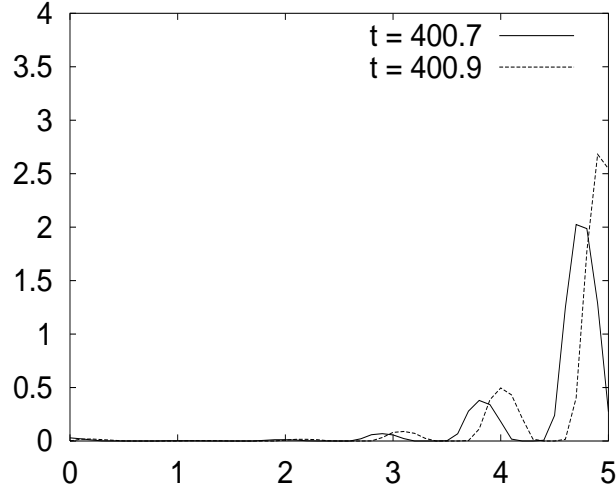}} \caption{\lab{fig5}Two snapshots of the maturation
profile for the numerical solution in figure \ref{fig4}. An exponentially growing travelling wave propagates
down the maturation axis.}
\end{center}
\end{figure}

The oscillations have period equal to the cell cycling time, equal to
one in our scaled model. A partial understanding of these oscillations
is afforded by the observation that if $h$ is constant and $M$ is
periodic with period $2\pi/\omega$, then
(\ref{4.6}) admits a solution
\be\lab{8.5.37}M=\sum_{p,q}c_{pq}e^{\sigma_q\xi+ip\omega (t-\xi)},\ee
provided $\sigma_q$ satisfies
\be\lab{8.5.38}\sigma=-A-G e^{-\sigma},\ee
where
\be\lab{8.5.39}A =bh, \quad G=-bh(1+\lambda).\ee
Since  $\lambda>0$, we have
$G+A<0$, and it is straightforward to show that there is always a
single positive root, which can be
labelled with $q=0$. The others are complex
(conjugates), and are labelled with increasing frequency as $q=\pm
1,\pm 2$, etc. Consideration of these complex roots then shows that
for small $|G|$, $\real\sigma_q <0$, so that the effect of the
oscillations dies away as the cells mature; this is what happens in
figure \ref{fig3}. However, for larger $|G|$, $\real\sigma_q >0$, and
the oscillations grow in amplitude as the cells mature. This causes
$\bar{M}$ to fluctuate, and thus also $h$, presumably entraining the
period of the oscillations to that of the delay. This description is
consistent with what is seen in figure \ref{fig5} (see also
figure \ref{fig6}). An approximate criterion for growth of periodic
perturbations with $\xi$ is when $G<-[A^2+\pi^2]^{1/2}$, i.\,e.,
\be\lab{4.7}\lambda\gsim\left[1+
\left(\frac{\pi}{bh}\right)^2\right]^{1/2}-1.\ee

\begin{figure}
\begin{center}
\resizebox{280pt}{200pt}{\includegraphics{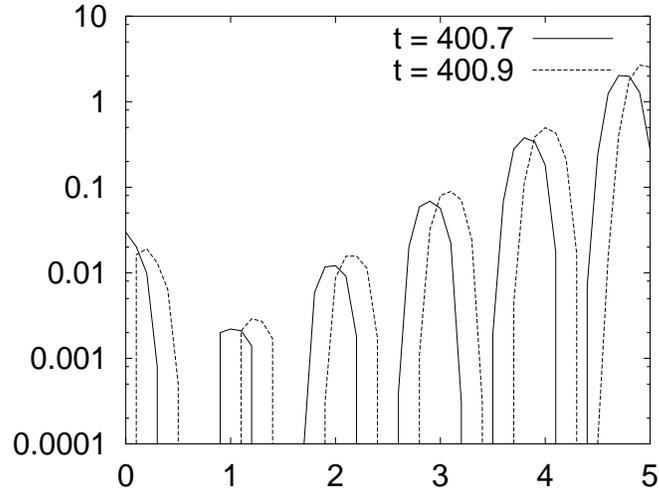}} \caption{\lab{fig6}The same graph as in figure
\ref{fig5}, except that a logarithmic scale is used. The exponential increase with $\xi$ is clearly seen.}
\end{center}
\end{figure}

\subsection*{Differentiation-controlled oscillations}

The final kind of oscillation that we see is induced by the peripheral
control of stem cell committal through the function $v_0(B)$. These
are essentially delay induced oscillations, where now the delay
involved is the maturation time. Because we suppose maturation time is
large, these are long period oscillations. They can be caused by
increasing the sensitivity of the peripheral controller, as shown in
figure \ref{fig7}.

\begin{figure}
\begin{center}
\resizebox{240pt}{200pt}{\includegraphics{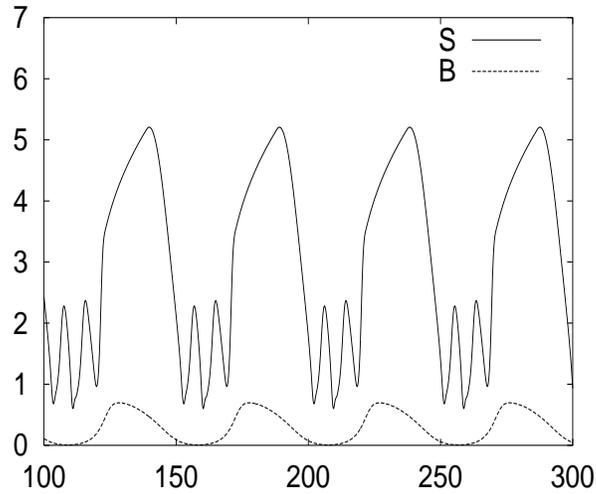}} \caption{\lab{fig7}Differentiation-controlled
oscillations due to enhancement of the peripheral controller function. Default parameters are used, except that
$v^*=v'=2$.}
\end{center}
\end{figure}

To understand the origin of these oscillations, let us suppose that
$\xi_f\gg 1$, or $\xi_F\gg
\tau$, meaning that the maturation time is significantly longer than
the cell cycle time, or equivalently that there are a large number of
generations in the cell lineage. Let us define
\be\lab{8.5.45}\varepsilon=\frac{1}{\xi_m},\ee
and the slow time and maturation scales
\be\lab{8.5.46}T=\varepsilon t,\quad X=\varepsilon\xi.\ee
We also define $\mu$ via
\be\lab{8.5.47}\lambda=\varepsilon\mu,\ee
and suppose that $\mu=O(1)$. Essentially we are revisiting the
relaxation oscillation analysis of Fowler and Mackey (2002). The partial
differential equation for $M$ takes the form
\be\lab{8.5.48}\pdone{M}{T}+\pdone{M}{X}=
\frac{b\left[h_{\varepsilon,\varepsilon}
M_{\varepsilon,\varepsilon}-hM\right]}{\varepsilon}+\mu
bh_{\varepsilon,\varepsilon} M_{\varepsilon,\varepsilon},\ee
and expanding in a Taylor series, we have
\be\lab{8.5.49}\pdone{[(1+bh)M]}{T}+\pdone{[(1+bh)M]}{X}\approx \mu bhM,\ee
with the boundary condition (taking $S=1$)
\be\lab{8.5.50}M=\nu v_0(B)\quad {\rm at}\quad X=0.\ee

If we suppose $h$ is constant (it is not, but it is not the dependence
of $h$ on $\bar{M}$ which causes the oscillations), then the solution
of this is
\be\lab{8.5.51}M=\nu v_0[B(T-X)]\exp\left[\frac{\mu bh
X}{1+bh}\right],\ee and the cell efflux at $X=1$ ($\xi=\xi_f$) is
\be\lab{8.5.52}M(1)=\nu av_0[B(T-1)],\ee
where the amplification factor $a$ is
\be\lab{8.5.53}a=\exp\left[\frac{\mu bh }{1+bh}\right].\ee
Therefore the blood cell  conservation law (\ref{3.15.2})
becomes the delay recruitment model
\be\lab{8.5.54}\varepsilon\delta \sdone{B}{T}=\nu av_0(B_1)- B.\ee
This is a standard delay recruitment equation with a unique steady
state. Oscillations will occur as a consequence of instability if
there are solutions $\sigma$ of (\ref{8.5.38}), i.\,e.,
\be\lab{4.8}\sigma=-A-Ge^{-\sigma},\ee
with positive real part. The values of $A$ and $G$ are
\be\lab{4.9}A=\frac{1}{\varepsilon\delta},\quad G=\frac{\nu
a|v_0^\prime |}{\varepsilon\delta}.\ee

The  equation (\ref{4.8}) is  well understood, see for
example Mackey (1978) or Murray (2002, pp.\ 23--26). It is a
transcendental equation with
an infinite number of  complex roots which
accumulate at the essential singularity at
$\sigma = -\infty$. It follows that the set of $\real\sigma$ is bounded
above. Consequently, there is an instability criterion which
determines when {\it
all} the roots $\sigma$ have negative real part, and this is  indicated
in  figure \ref{fig8.8}.
The three curves in the figure are given by $G=-A$,
$G=\exp[-(1+A)]$, and the Hopf bifurcation curve $G=G_0(A)$,
which is given parametrically by
\be\lab{15}A=-\frac{\Omega}{\tan\Omega},\quad
G_0=\frac{\Omega}{\sin\Omega},\ee
where $\Omega\in [0,\pi]$. Since $G$ and $A$ are positive, oscillatory
instability occurs precisely if $G>G_0(A)$.
Since $G_0\sim A$ as $A\to\infty$,
the instability criterion for large $A$ is
simply $G\gsim A$, i.\,e.,
\be\lab{4.10}\nu|v_0^\prime |\exp\left[\frac{\mu bh }{1+bh}\right]>1.\ee
Instability is promoted by increasing $|v_0^\prime |$, for example, as
indicated in figure \ref{fig7}.

\begin{figure}
\begin{center}
\resizebox{280pt}{220pt}{\includegraphics{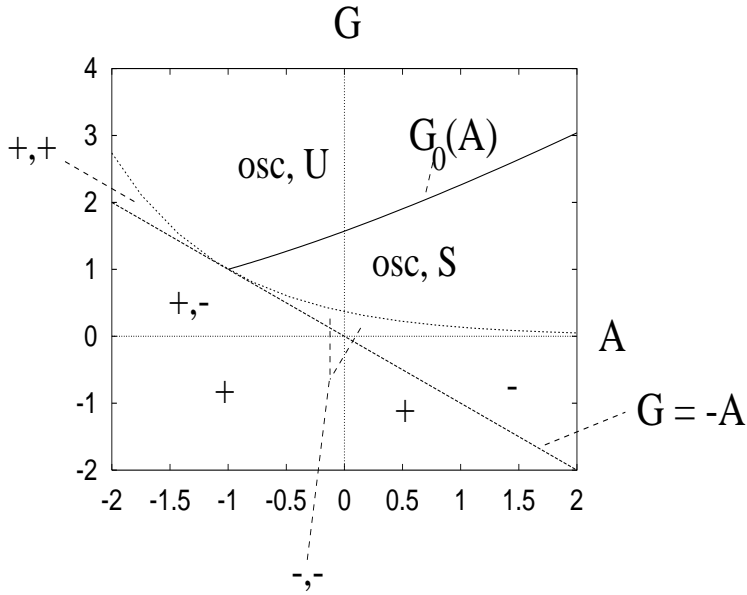}} \caption{\lab{fig8.8}Stability map for (\ref{4.8}). The
plus and minus signs indicate the sign of real values of the growth rate $\sigma$, when these exist. A Hopf
bifurcation occurs as $G$ increases through $G_0(A)$, and $G_0\approx A$ for large $A$.}
\end{center}
\end{figure}

\section{Conclusions}
\reseteqnos{5}

In this paper we have studied the onset of oscillations in a model
of blood cell production which includes a description of cell
cycling and proliferation, and also of differentiation and
maturation. The model formulation extends the work of previous
authors by correcting an apparent  inconsistency in the
description of the primitive stem cell population, and also by
including the simultaneous control of stem cell proliferation,
stem cell committal, and blast cell proliferation. All three
controls can cause oscillations for appropriate values of control
parameters.

Previous results concerning stem cell oscillations are reproduced
(see figure \ref{fig3}) but these oscillations are harder to
obtain when the parameter $\nu$ is small, and in addition they
hardly affect the mature blood cell population, without additional
destabilisation of the blast cell proliferation. The reason for
this is that an $O(1)$ oscillation in the stem cell population
only causes an $O(\nu)$ oscillation in the blast cell committal
rate, and this amplitude propagates through the differentiating
cells. Thus one consequence of stem cell paucity is that any
instability in the stem cell population itself is hardly
manifested in the blood cell production. From the point of view of
survival and  control, this is of course a positive result.

Instability in the proliferation of blast cells due to enhancement
of the proliferative controller $h(\bar{M})$ causes oscillations
which propagate down the maturation axis, and are amplified as
they progress. The result of this is shown in figures \ref{fig4},
\ref{fig5} and \ref{fig6}. The oscillations have a period equal to
the cell cycling time. The mechanism for these oscillations
appears to be a destabilisation of the maturing cell
amplification, together with a type  of resonance which ties the
period to the delay.

The final kind of oscillation is induced by  enhanced peripheral
control, as seen in figure \ref{fig7}. Stem cell paucity implies
that $\nu\ll 1$, and consequently that $\xi_f\gg 1$, and thus that
the oscillation period (controlled by the delay $\xi_f$) is long.
This allows an approximate reduction of the partial differential
delay equation to a simple first order differential delay
equation, which is simply analysed. In particular, if a threshold
form of peripheral controller is used, blood cell counts can
decrease to zero, as can be the case in cyclical neutropenia.

Finally, and as shown in figure \ref{fig8}, a combination of all
three destabilising mechanisms can lead to oscillations which
operate on both the slow, peripherally controlled time scale and
the fast, proliferatively controlled one. We consider this
observation to be a possible explanation of the apparent fact in
figure \ref{figcn} that both
reticulocytes and platelets appear to oscillate on  a fast as  well as
a slow time scale. Further study
of this behaviour requires the extension of this model to
accommodate multiple cell lineages.

\begin{figure}
\begin{center}
\resizebox{240pt}{200pt}{\includegraphics{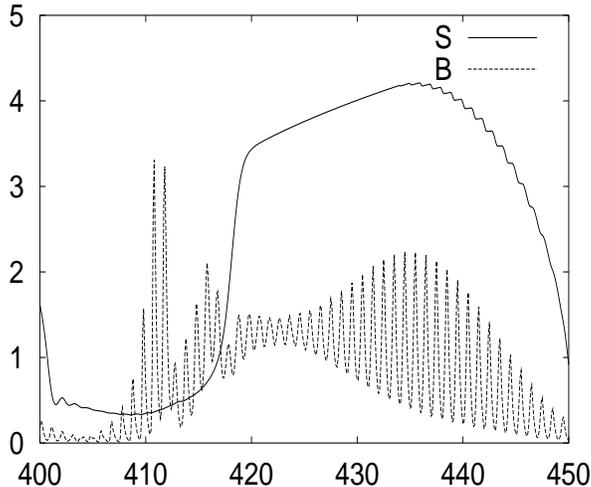}} \caption{\lab{fig8}The effect  of switching on all three
instability mechanisms. Default parameters are used, except that $v^*=v'=2$, $\lambda_0=0.05$, and
$\lambda=0.6$. }
\end{center}
\end{figure}

\section*{References}

\begin{description}

\item  Bernard, S., J.\ B\'elair and M.\,C.\ Mackey 2003 Oscillations
in cyclical
neutropenia: new evidence based on mathematical modeling. J.\
Theor.\ Biol.\ {\bf 223}, 283--298.

\item Chikkappa, G.,  G.\ Borner, H.\ Burlington,
          A.\,D.\ Chanana, E.\,P.\ Cronkite, S.\ \"{O}hl,
  M.\ Pavelec  and  J.\,S.\ Robertson 1976  Periodic oscillation
  of blood leukocytes, platelets, and
  reticulocytes in a patient with chronic myelocytic leukemia.
Blood {\bf 47}, 1023--1030.

\item Crabb, R., M.\,C.\ Mackey and A.\,D.\ Rey 1996 Propagating
fronts, chaos and  multistability in a cell replication model.  Chaos
{\bf 6}, 477--492.

\item Dale, D.\,C.\ and W.\,P.\ Hammond IV 1988 Cyclical neutropenia:
  a clinical review. Blood Revs.\ {\bf 2}, 178--185.

\item Dyson, J., R.\ Villella-Bresson and G.\,F.\ Webb 1996 A singular
transport equation modelling a proliferating maturity structured cell
population.  Can.\ Appl.\ Math.\ Quart.\ {\bf 4}, 65--95.

\item Fokas, A.\,S., J.\,B.\ Keller and B.\,D.\ Clarkson 1991
Mathematical model of granulocytopoiesis and chronic myelogenous
leukaemia. Cancer Research {\bf 51}, 2084--2091.

\item  Fortin, P.\  and  M.\,C.\ Mackey 1999 Periodic chronic myelogenous
leukemia: Spectral analysis of blood cell counts and etiological
implications. Br. J. Haematol., {\bf 104}, 336--345.

\item Fowler, A.\,C.\ and M.\,C.\ Mackey 2002 Relaxation oscillations in
a class of delay differential equations.  SIAM J.\ Appl.\ Math.\ {\bf
  63},
299--323.

\item Guerry IV, D., D.\,C.\ Dale, M.\ Omine, S.\ Perry and S.\,M.\
  Wolff 1973 Periodic hematopoiesis in human cyclic neutropenia. J.\
  Clin.\ Investig. {\bf 52}, 3,220--3,230.

\item Haurie, C.,  D.\,C.\ Dale and  M.\,C.\ Mackey 1998 Cyclical
neutropenia and other
periodic hematological disorders: a review of mechanisms and mathematical
models. Blood {\bf 92}, 2629--2640.

\item Mackey, M.\,C.\ 2001 Cell kinetic status of haematopoietic
stem cells. Cell. Prolif. {\bf 34}, 71--83.

\item Mackey, M.\,C.\ 1978 A unified hypothesis for the origin of
aplastic anaemia and periodic haematopoesis.  Blood {\bf 51},
941--956.

\item Mackey, M.\,C.\ and R.\ Rudnicki 1994 Global stability in a
delayed partial differential equation describing cellular replication.
J.\ Math.\ Biol.\ {\bf 33}, 89--109.

\item Mackey, M.\,C.\ and R.\ Rudnicki 1999 A new criterion for the
global stability of simultaneous cell replication and maturation
processes.  J.\ Math.\ Biol.\ {\bf 38}, 195--219.

\item Murray, J.\,D.\ 2002 Mathematical biology. I: an
introduction. Springer-Verlag, Berlin.

\item Rey, A.\,D.\ and M.\,C.\ Mackey 1993 Multistability and boundary
layer development in a transport equation with delayed arguments.
Can.\ Appl.\ Math.\ Quart.\ {\bf 1}, 61--81.

\end{description}

\end{document}